\documentclass[a4paper,11pt]{article}
\usepackage{pos}

\usepackage[utf8]{inputenc} 
\usepackage{amsmath}
\usepackage{bm}
\usepackage{multirow}
\usepackage{soul}
\hypersetup{colorlinks=true, urlcolor=cyan, linkcolor=cyan}

\title{Constraining new physics from Higgs measurements\\ with Lilith-2}
\ShortTitle{Lilith-2}

\author[a,b]{Marius Bertrand}
\author*[a]{Sabine Kraml}
\author[c]{Tran Quang Loc}
\author[d]{Dao Thi Nhung}
\author[d]{Le Duc Ninh}

\affiliation[a]{Laboratoire de Physique Subatomique et de Cosmologie, Universit\'e Grenoble-Alpes, CNRS/IN2P3,\\ 53 Avenue des Martyrs, F-38026 Grenoble, France}

\affiliation[b]{Facult\'e des Sciences d’Orsay, Univ.\ Paris-Saclay, 15 rue Georges Clemenceau, F-91405 Orsay, France}

\affiliation[c]{Department of Applied Mathematics \& Theoretical Physics, University of Cambridge, Wilberforce Road, Cambridge CB3 0WA, United Kingdom}

\affiliation[d]{Institute for Interdisciplinary Research in Science and Education, ICISE, 55121 Quy Nhon, Vietnam}

\emailAdd{gusbertrand@orange.fr}
\emailAdd{sabine.kraml@lpsc.in2p3.fr}
\emailAdd{loctranq@gmail.com}
\emailAdd{dtnhung@ifirse.icise.vn}
\emailAdd{ldninh@ifirse.icise.vn}

\abstract{{\tt Lilith} is a public Python library for constraining new physics from Higgs signal strength measurements. Version 2.0 of Lilith comes with an extensive XML database which includes the ATLAS and CMS Run 2 Higgs results for 36/fb, in addition the the Run~1 results. Both the code and the database were extended from the ordinary Gaussian approximation employed in Lilith-1.1 to using variable Gaussian and Poisson likelihoods. Moreover, Lilith-2 can make use of correlation matrices of arbitrary dimension. We will report on these novelties and ongoing developments. The importance of how correlations and uncertainties are treated will be demonstrated by means of detailed validations of the implemented experimental results. Moreover, we show the effects for global fits of reduced Higgs couplings, 2HDMs of Type~I and Type~II, and invisible Higgs decays.
The program is publicly available on GitHub and can be used to constrain a wide class of new physics scenarios.}

\FullConference{%
  Tools for High Energy Physics and Cosmology - TOOLS2020\\
  2-6 November, 2020\\
  Institut de Physique des 2 Infinis (IP2I), Lyon, France}

\begin{document}
\maketitle

%%%%%%%%%%%%%%%%%%%%%%%%%%%%%%%%%%%%%%%%%%%%%%%%%%%%%%
\section{Introduction}
%%%%%%%%%%%%%%%%%%%%%%%%%%%%%%%%%%%%%%%%%%%%%%%%%%%%%%

The LHC runs in 2010--2012 and 2015--2018 have led to a wealth of experimental results on the 125~GeV Higgs boson. 
From this emerges an increasingly precise picture of the various Higgs production and decay processes, and consequently 
of the Higgs couplings to the other particles of the Standard Model (SM), notably gauge bosons and third generation fermions. 
With all measurements so far agreeing with SM predictions, this poses severe constraints on scenarios of new physics,  
in which the properties of the observed Higgs boson could be affected in a variety of ways. 
Indeed, beyond-the-SM (BSM) theories typically predict deviations from the SM expectations, due to mixing effects in enlarged Higgs sectors, loop contributions from new particles (or higher-order effects in effective field theories), new decay modes, and so on. The Higgs data have therefore become a major guideline for BSM theories. 

Assessing the compatibility of a non-SM Higgs sector with the ATLAS and CMS results requires to construct a likelihood, 
which is a non-trivial task. While this is best done by the experimental collaborations themselves, having at least an approximate global likelihood is very useful, as it allows theorists to pursue in-depth studies of the implications for their models. 
For this reason, the public code {\tt Lilith}~\cite{Bernon:2014vta,Bernon:2015hsa,Kraml:2019sis} was created, making use of the Higgs signal strength measurements published by ATLAS and CMS, and the Tevatron experiments. 
%\footnote{For a discussion of the use and usability of signal strength results, and recommendations on their presentation, see~\cite{Boudjema:2013qla}.}  
%
{\tt Lilith} is a light-weight Python library that uses as a primary input signal strength results
\begin{equation}
	\mu(X,Y) \equiv \frac{\sigma(X)\,{\rm BR}(H\to Y)}{\sigma^{\rm SM}(X)\,{\rm BR}^{\rm SM}(H\to Y)} \,,
	\label{eq:signalstr}
\end{equation}
in which the fundamental production and decay modes are unfolded from experimental categories. 
Here, the main production mechanisms $X$ are: gluon fusion ($\rm ggH$), vector-boson fusion ($\rm VBF$), associated production with an electroweak gauge boson ($\rm WH$ and $\rm ZH$, collectively denoted as $\rm VH$) and associated production with top quarks, mainly $\rm ttH$ but also $\rm tH$. The main decay modes $Y$ accessible at the LHC are $H \to \gamma\gamma$, $H \to Z Z^* \to 4\ell$ and $H \to W W^* \to 2\ell2\nu$ (with $\ell \equiv e,\mu$), $H \to b\bar b$, $H \to \tau^+\tau^-$ and recently also $H \to \mu^+\mu^-$.

Version~2 of {\tt Lilith}~\cite{Kraml:2019sis} features a better treatment of asymmetric uncertainties through the use of variable Gaussian and Poisson likelihoods (as compared to the normal Gaussian approximation in version 1).  Moreover, {\tt Lilith-2} can make use of correlation matrices of arbitrary dimension. The experimental results used are stored in an easily extendible XML database. 
The current official release comes with database DB~19.09, which includes the full set of published  ATLAS and CMS Run~2 Higgs results for 36~fb$^{-1}$ as of September 2019. 
An update to the available full Run~2 luminosity  ($\approx140$~fb$^{-1}$) results is in progress.

%%%%%%%%%%%%%%%%%%%%%%%%%%%%%%%%%%%%%%%%%%%%%%%%%%%%%%
\section{The signal strength framework}
%%%%%%%%%%%%%%%%%%%%%%%%%%%%%%%%%%%%%%%%%%%%%%%%%%%%%%

The signal strength framework is based on the narrow-width approximation and on the assumption that new physics results only in the scaling of SM Higgs processes. 
The likelihood in terms of $\mu(X,Y)$, Eq.~\eqref{eq:signalstr}, allows for reinterpretation of the results within models  
where the signal acceptances for the $(X,Y)$ are to good approximation the same as in the SM; this applies to 
models which i) have the same tensor structure as the SM and ii)~have no new production modes; see e.g.\ the discussion in~\cite{Boudjema:2013qla}. 
This framework can be used to constrain a wide variety of new physics models, see for example \cite{Belanger:2013xza} and references therein.

Results given in terms of signal strengths can be matched to new physics scenarios with the introduction of factors ${\bm C}_X$ and ${\bm C}_Y$ that scale the amplitudes for the production and decay of the SM Higgs boson, respectively, as 
\begin{equation}
   \mu(X,Y) = \frac{{\bm C}_X^2 {\bm C}_Y^2}{\sum_Y {\bm C}_Y^2\, {\rm BR}^{\rm SM}(H\to Y)} 
   \label{eq:signalstr2}
\end{equation}
for the different production modes $X\in({\rm ggH},\, {\rm VBF},\, {\rm WH},\, {\rm ZH},\, {\rm ttH},\, \ldots)$ and 
decay modes $Y\in(\gamma\gamma$, $ZZ^*$, $WW^*$, $b\bar{b}$, $\tau\tau$, $\ldots)$, 
where the sum runs over all decays that exist for the SM Higgs boson. 
The factors ${\bm C}_X$ and ${\bm C}_Y$ can be identified to (or derived from) reduced couplings appearing in an effective Lagrangian.%
\footnote{See, e.g., \cite{Bernon:2015hsa} for   detailed explanations.}  
Following~\cite{Belanger:2012gc,Belanger:2013xza} and subsequent publications, we employ the notation
\begin{equation}
	\label{eq:lagrangian}
	\mathcal{L} = g \left[   C_W m_W W^\mu W_\mu + C_Z \frac{m_Z}{\cos \theta_W} Z^\mu Z_\mu 
	                                     - \sum_f C_f \frac{m_f}{2m_W} f\bar{f}      \right] H \, ,
\end{equation}
where $C_{W,Z}$ and $C_f$ ($f=t,b,c,\tau,\mu$) are bosonic and fermionic reduced couplings, respectively. 
In the limit where all reduced couplings go to 1, the SM case is recovered. 
In  addition  to  these  tree-level  couplings,  we  define  the  loop-induced couplings $C_g$ and $C_\gamma$ of the Higgs to 
gluons and photons, respectively. If no new particles appear in the loops, $C_g$ and $C_\gamma$ are computed 
from the couplings in Eq.~\eqref{eq:lagrangian} following the procedure established in~\cite{Heinemeyer:2013tqa}. 
Alternatively, $C_g$ and $C_\gamma$  can be taken as free parameters. 
Apart from the different notation, this is equivalent to the so-called ``kappa framework'' of \cite{Heinemeyer:2013tqa}. 
Finally note that often a subset of the $C$'s in Eq.~\eqref{eq:lagrangian} is taken as universal, 
in particular $C_V\equiv C_W=C_Z$ (custodial symmetry), 
$C_U\equiv C_t=C_c$ and $C_D\equiv C_b=C_\tau=C_\mu$ like in the Two-Higgs-doublet model (2HDM) of Type~II, 
or $C_F\equiv C_U=C_D$ as in the 2HDM of Type~I. 

Last but not least, while the signal strength framework in principle requires that the Higgs signal be a sum of processes 
that exist for the SM Higgs boson, decays into invisible or undetected new particles, 
affecting only the Higgs total width, can be accounted for through 
\begin{equation}
   \mu(X,Y) \to \left[ 1 - {\rm BR}(H\to {\rm inv.}) - {\rm BR}(H\to {\rm undet.}) \right]\,\mu(X,Y) 
\label{eq:BRinv}
\end{equation}
without spoiling the approximation.

%%%%%%%%%%%%%%%%%%%%%%%%%%%%%%%%%%%%%%%%%%%%%%%%%%%%%%
\section{Lilith 2.0 release, database 19.09}
%%%%%%%%%%%%%%%%%%%%%%%%%%%%%%%%%%%%%%%%%%%%%%%%%%%%%%

\subsection*{Likelihood approximations}

For a proper inclusion of the recent Run~2 results from ATLAS and CMS, several improvements were necessary in {\tt Lilith}. 
First of all, the ordinary Gaussian approximation was no longer sufficient to model the likelihood.   
We have therefore extended the parametrisation of the likelihood to Gaussian functions of variable width (``variable Gaussian'') as well as generalised Poisson functions, following the prescriptions of \cite{Barlow:2004wg}.
Moreover, {\tt Lilith} can now make use of correlation matrices of arbitrary dimension (for Gaussian likelihoods; Poisson likelihoods are still limited to 2 dimensions). 
We have also added the $\rm tH$ and the gluon-initiated ${\rm ZH}$ production modes, and corrected some minor bugs in the code. The exact way all the above is implemented is described in detail in  \cite{Kraml:2019sis}. 

\begin{figure}[t]\centering
\includegraphics[width=0.45\textwidth]{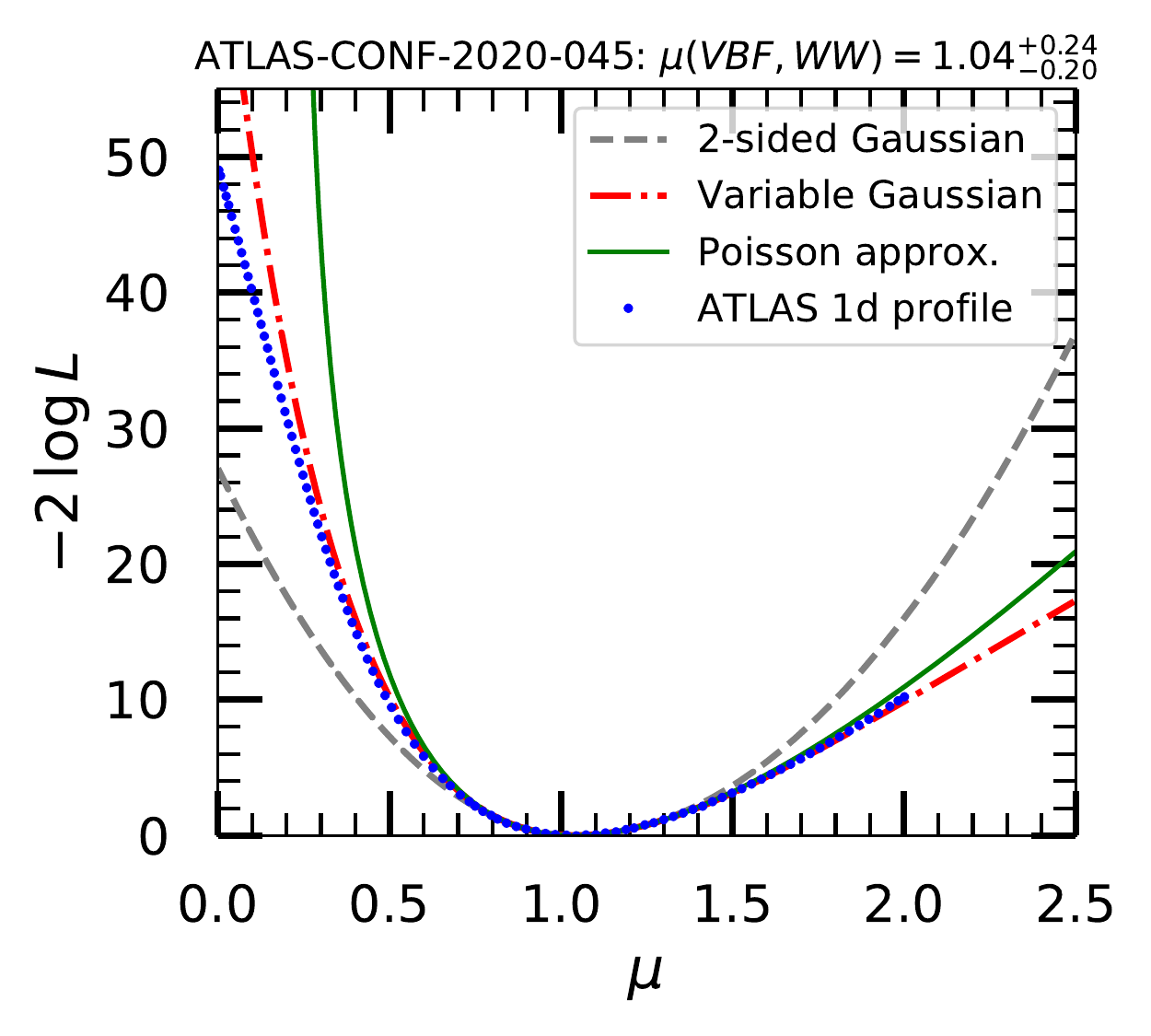}\hspace{-12mm}%
\vspace*{-2mm}
\caption{Negative log-likelihoods constructed from $\mu({\textrm{VBF}},\,WW)=1.04^{+0.24}_{-0.20}$~\cite{ATLAS-CONF-2020-045} using 2-sided Gaussian (grey dashed), variable Gaussian (red dash-dot) and Poisson (green solid) approximations. For comparison, the 1D profile likelihood determined by ATLAS is shown as blue dotted line.}
\label{fig:LogLikeComparison}
\end{figure}

The importance of appropriately modelling the \emph{shape} of the likelihood is illustrated in Figure~\ref{fig:LogLikeComparison} by means of the ATLAS VBF$\to H\to WW^*\to 2\ell 2\nu$ measurement \cite{ATLAS-CONF-2020-045}. This ATLAS publication reports $\mu({\textrm{VBF}},\,WW)=1.04^{+0.24}_{-0.20}$ as well as a plot of the likelihood profile, which we digitised. As can be seen, the ordinary 2-sided Gaussian, variable Gaussian, and Poisson approximations reproduce the actual likelihood profile, $-2\log L(\mu)$, with different accuracy. In this example, the best-performing method is the variable Gaussian, but there are other cases where the Poisson distribution gives a better result. 
 
\subsection*{Experimental results in DB19.09}

\begin{figure}[t]\centering
\includegraphics[width=0.43\textwidth]{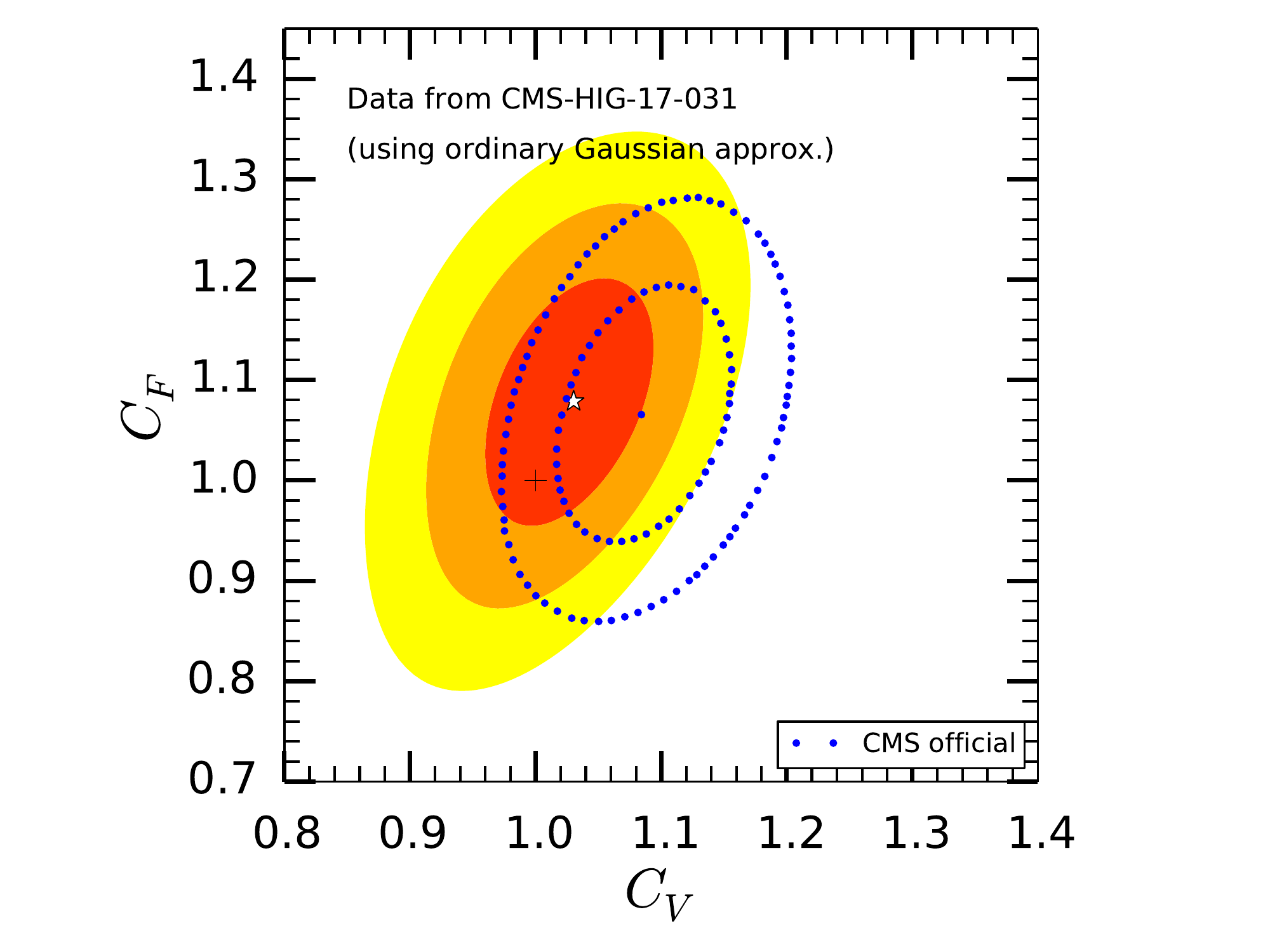}\hspace{-10mm}%
\includegraphics[width=0.45\textwidth]{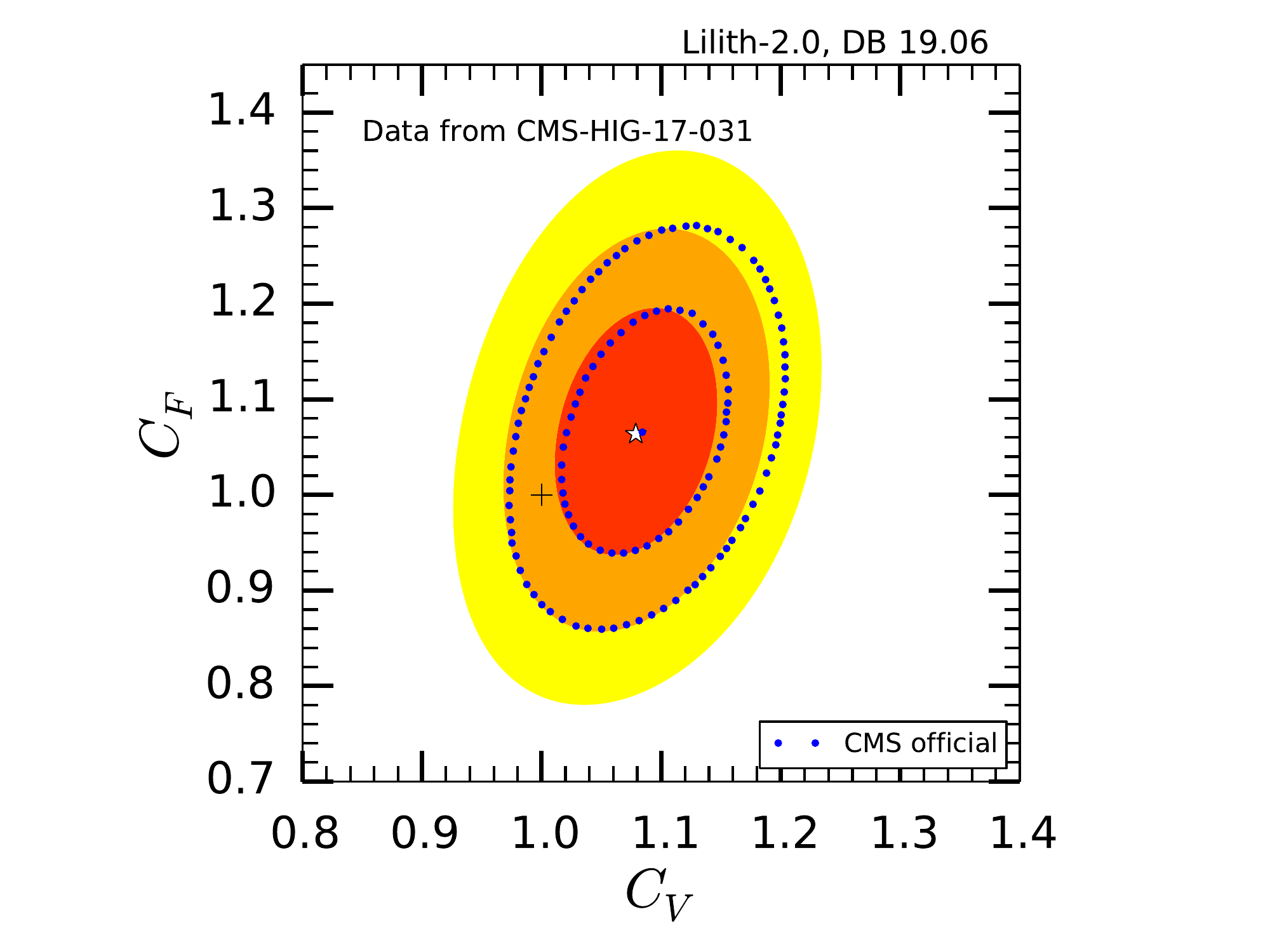}%
\vspace*{-2mm}
\caption{Fit of $C_F$ vs.\ $C_V$ using best-fit values and uncertainties for the signal strengths for each production (ggH, VBF, WH, ZH, ttH) and decay ($\gamma\gamma$, $ZZ$, $WW$, $\tau\tau$, $b\bar b$, $\mu\mu$) mode combination together with the $24\times 24$ correlation matrix from the CMS combination paper~\cite{Sirunyan:2018koj}. 
The  $1\sigma$,  $2\sigma$ and $3\sigma$ regions are shown as red, orange and yellow areas, respectively, 
and compared to the $1\sigma$ and $2\sigma$ contours from CMS (blue dots). On the left, the plus and minus uncertainties are described by ordinary Gaussian functions, while on the right they are described by Gaussians of variable width.}
\label{fig:CMSvalidation}
\end{figure}

The current official release, {\tt Lilith-2.0}, %is written in {\tt python2} and available from 
%\begin{quote}
%\url{https://github.com/sabinekraml/Lilith-2/releases}. 
%\end{quote}
%It 
comes with database DB~19.09, which includes the full set of published  ATLAS and CMS Run~2 Higgs results for 36~fb$^{-1}$ as of September 2019, 
summarised in Tables~\ref{tab:ATLASresults} and \ref{tab:CMSresults}. %, respectively.  
Each result has been carefully validated. 
As an example, Figure~\ref{fig:CMSvalidation} shows the validation %of the implementation of the data from 
for the CMS combination paper~\cite{Sirunyan:2018koj} by means of a $C_F$ vs.\ $C_V$ fit. Both versions of the implementation 
use the full $24\times 24$ correlation matrix given in~\cite{Sirunyan:2018koj}. However, in the version shown on the left, the plus and minus uncertainties for the 24 channels are described by ordinary Gaussian functions with symmetrised errors, while on the right they are described by Gaussians of variable width. As can be seen, there is a small difference in the resulting errors on $C_F$ and  $C_V$, but most importantly there is a significant shift in the best-fit values.  Clearly, the variable Gaussian agrees much better with the official CMS fit, and is hence used in the {\tt Lilith} database.

\begin{figure}[t]\centering
\mbox{\hspace{-3mm}\includegraphics[width=0.4\textwidth]{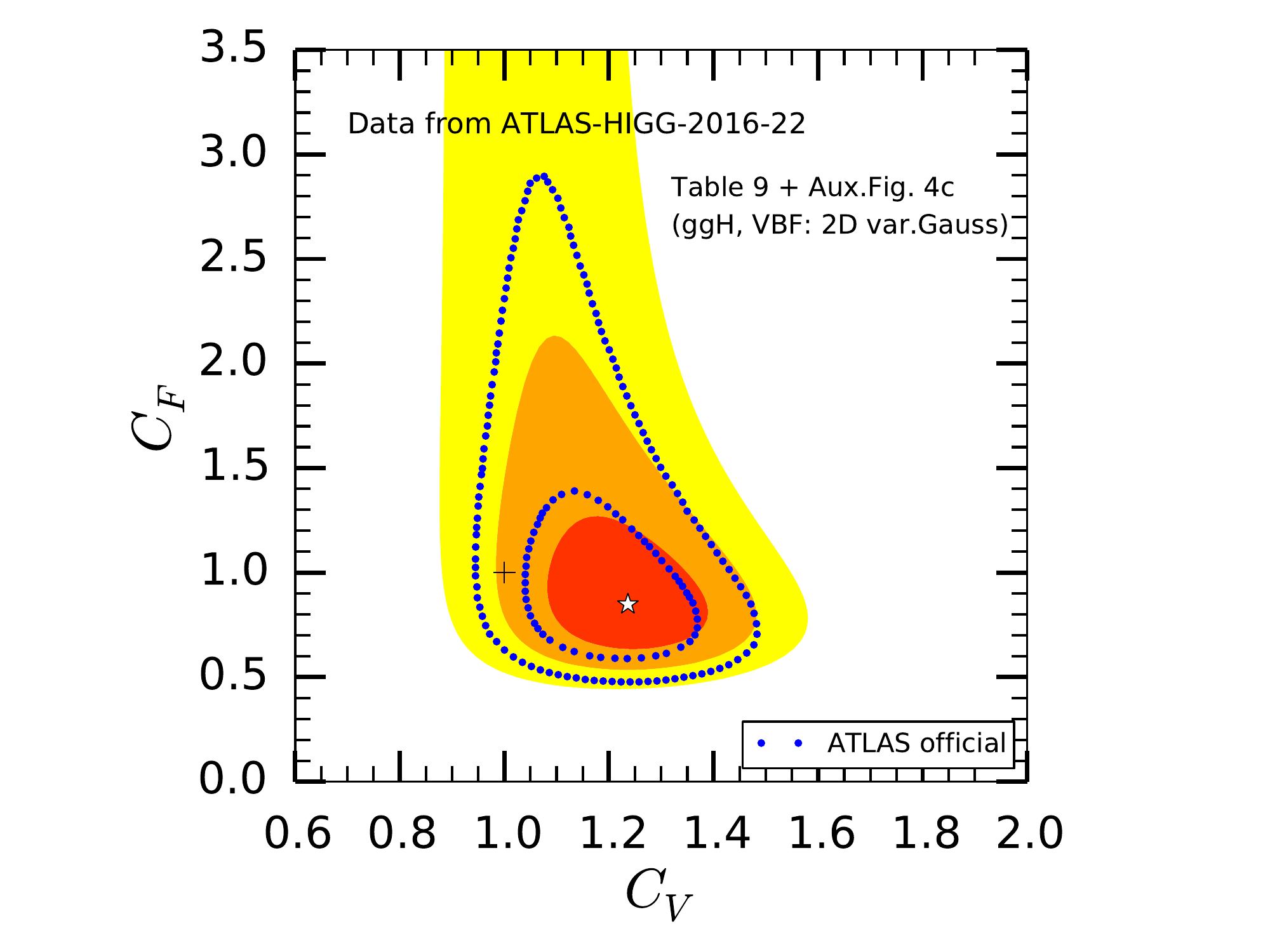}\hspace{-11mm}%
\includegraphics[width=0.4\textwidth]{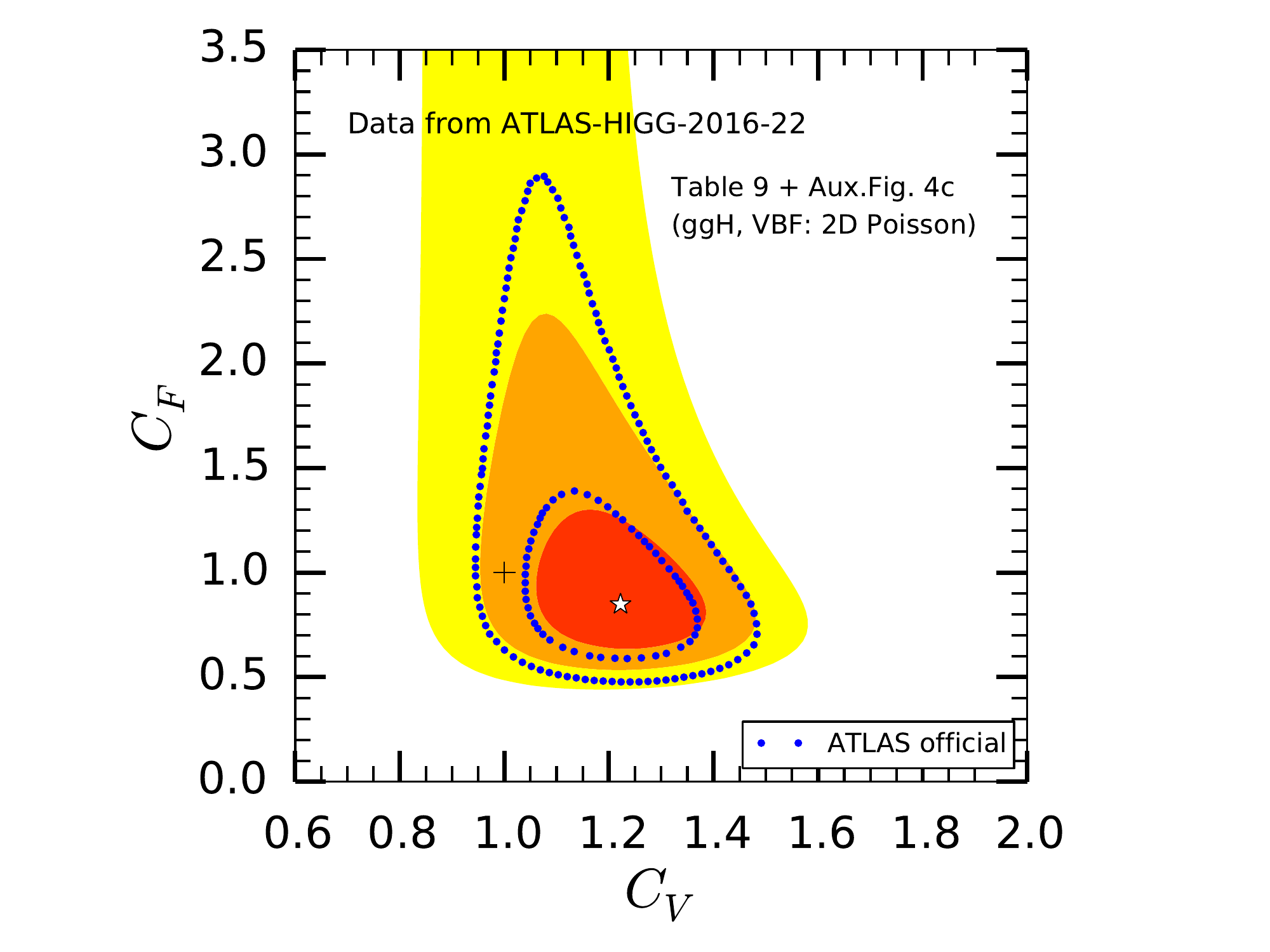}\hspace{-11mm}%
\includegraphics[width=0.4\textwidth]{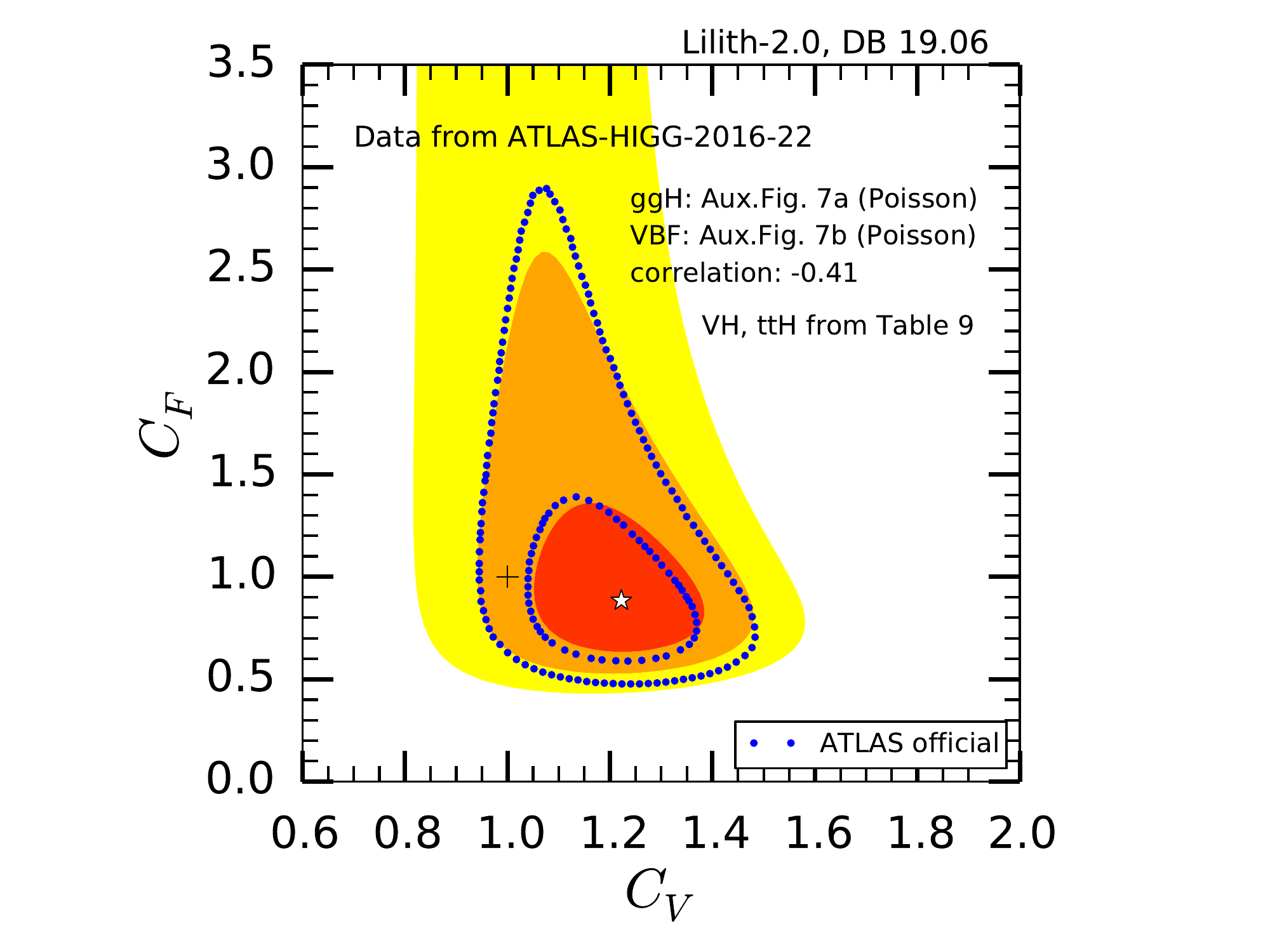}}
\vspace*{-4mm}
\caption{Fit of $C_F$ vs.\ $C_V$ for data from the ATLAS $H\to ZZ^*$ analysis, trying different implementations 
of the $\mu({\rm ggH},\,ZZ^*)$ and $\mu({\rm VBF},\,ZZ^*)$ values given by ATLAS; see text for details. The $68\%$,  $95\%$ and $99.7\%$~CL regions obtained with {\tt Lilith} are shown as red, orange and yellow areas, 
and compared to the $68\%$ and  $95\%$~CL contours from ATLAS (in blue).
The best-fit point from {\tt Lilith} is marked as a white star and the SM as a $+$.}
\label{fig:ATLASvalidation}
\end{figure}

In some cases, reproducing the official coupling fits is more tricky. As an illustrative example, we show in Figure~\ref{fig:ATLASvalidation} fits of $C_F$ vs.\ $C_V$ for the data from ATLAS-HIGG-2016-22 ($H\to ZZ^*$) \cite{Aaboud:2017vzb}. The left and middle plots  use $\mu({\rm ggH},\,ZZ^*)=1.11^{+0.23}_{-0.21}$ and $\mu({\rm VBF},\,ZZ^*)=4.0^{+1.75}_{-1.46}$ from Table~9 of \cite{Aaboud:2017vzb} with correlation $\rho=-0.41$ from Aux.~Fig.~4c on the analysis twiki page. 
In the left plot, the likelihood of $\mu({\rm ggH},\,ZZ^*)$ vs.\ $\mu({\rm VBF},\,ZZ^*)$ is parametrised as a 
2D variable Gaussian, in the middle plot as a 2D Poisson distribution. Neither gives a very good agreement with the official ATLAS coupling fit. Fortunately, ATLAS also provided 1D profile likelihood plots in their auxiliary material. Fitting these curves as Poisson distributions, we obtain $\mu({\rm ggH},\,ZZ^*)\simeq1.12^{+0.25}_{-0.22}$ and $\mu({\rm VBF},\,ZZ^*)\simeq3.88^{+1.75}_{-1.46}$. With these values, the official ATLAS fit of $C_F$ vs.\ $C_V$ can be reproduced well, see the right plot in Figure~\ref{fig:ATLASvalidation}.

\begin{table}[t!]\centering
\begin{tabular}{l | cccccccc}
mode & $\gamma\gamma$ & $ZZ^*$ & $WW^*$ & $\tau\tau$ & $b\bar b$ & $\mu\mu$ & inv. \\
\hline
ggH & \cite{Aaboud:2018xdt} & \cite{Aaboud:2017vzb} & \cite{Aaboud:2018jqu} & \cite{Aaboud:2018pen} & -- & \cite{Aaboud:2017ojs} & --\\
VBF &  \cite{Aaboud:2018xdt} & \cite{Aaboud:2017vzb} & \cite{Aaboud:2018jqu} & \cite{Aaboud:2018pen} & \cite{Aaboud:2018gay} & \cite{Aaboud:2017ojs} & \cite{Aaboud:2019rtt} \\
WH & \multirow{2}{*}{\cite{Aaboud:2018xdt}} & \multirow{2}{*}{\cite{Aaboud:2017vzb}} & \cite{Aad:2019lpq} & -- & \cite{Aaboud:2017xsd} & 
\multirow{2}{*}{\cite{Aaboud:2017ojs}} & -- \\
ZH &  &  & \cite{Aad:2019lpq} & -- & \cite{Aaboud:2017xsd} &  & \cite{Aaboud:2017bja} \\
ttH & \cite{Aaboud:2018xdt} & \cite{Aaboud:2017vzb,Aaboud:2017jvq} & \cite{Aaboud:2017jvq} & \cite{Aaboud:2017jvq} & \cite{Aaboud:2017jvq,Aaboud:2017rss} & -- & -- \\ 
\end{tabular}
\caption{Overview of ATLAS Run~2 results included in DB\,19.09.} 
\label{tab:ATLASresults}
\end{table}

\begin{table}[t!]\centering
\begin{tabular}{l | ccccccc}
mode & $\gamma\gamma$ & $ZZ^*$ & $WW^*$ & $\tau\tau$ & $b\bar b$ & $\mu\mu$ & inv. \\
\hline
ggH & \cite{Sirunyan:2018koj} & \cite{Sirunyan:2018koj} & \cite{Sirunyan:2018koj} & \cite{Sirunyan:2018koj} & \cite{Sirunyan:2018koj} & \cite{Sirunyan:2018koj} & \cite{Sirunyan:2018owy} \\
VBF &  \cite{Sirunyan:2018koj} & \cite{Sirunyan:2018koj} & \cite{Sirunyan:2018koj} & \cite{Sirunyan:2018koj} &-- & \cite{Sirunyan:2018koj} & \cite{Sirunyan:2018owy} \\
WH &  \cite{Sirunyan:2018koj} & \cite{Sirunyan:2018koj} & \cite{Sirunyan:2018koj} & \cite{Sirunyan:2018cpi} & \cite{Sirunyan:2018koj} & -- & \cite{Sirunyan:2018owy} \\
ZH & \cite{Sirunyan:2018koj} & \cite{Sirunyan:2018koj} & \cite{Sirunyan:2018koj} & \cite{Sirunyan:2018cpi} & \cite{Sirunyan:2018koj} & -- & \cite{Sirunyan:2018owy} \\
ttH & \cite{Sirunyan:2018koj} & \cite{Sirunyan:2018koj} & \cite{Sirunyan:2018koj} & \cite{Sirunyan:2018koj} & \cite{Sirunyan:2018koj} & -- & -- \\
\end{tabular}
\caption{Overview of CMS Run~2 results included in DB\,19.09. Note that we use the full $24\times 24$ correlation matrix 
for the signal strengths for each production and decay mode combination provided in \cite{Sirunyan:2018koj}.}
\label{tab:CMSresults}
\end{table}

\subsection*{Combined coupling fits}

\begin{figure}[t!]\centering
\includegraphics[width=0.4\textwidth]{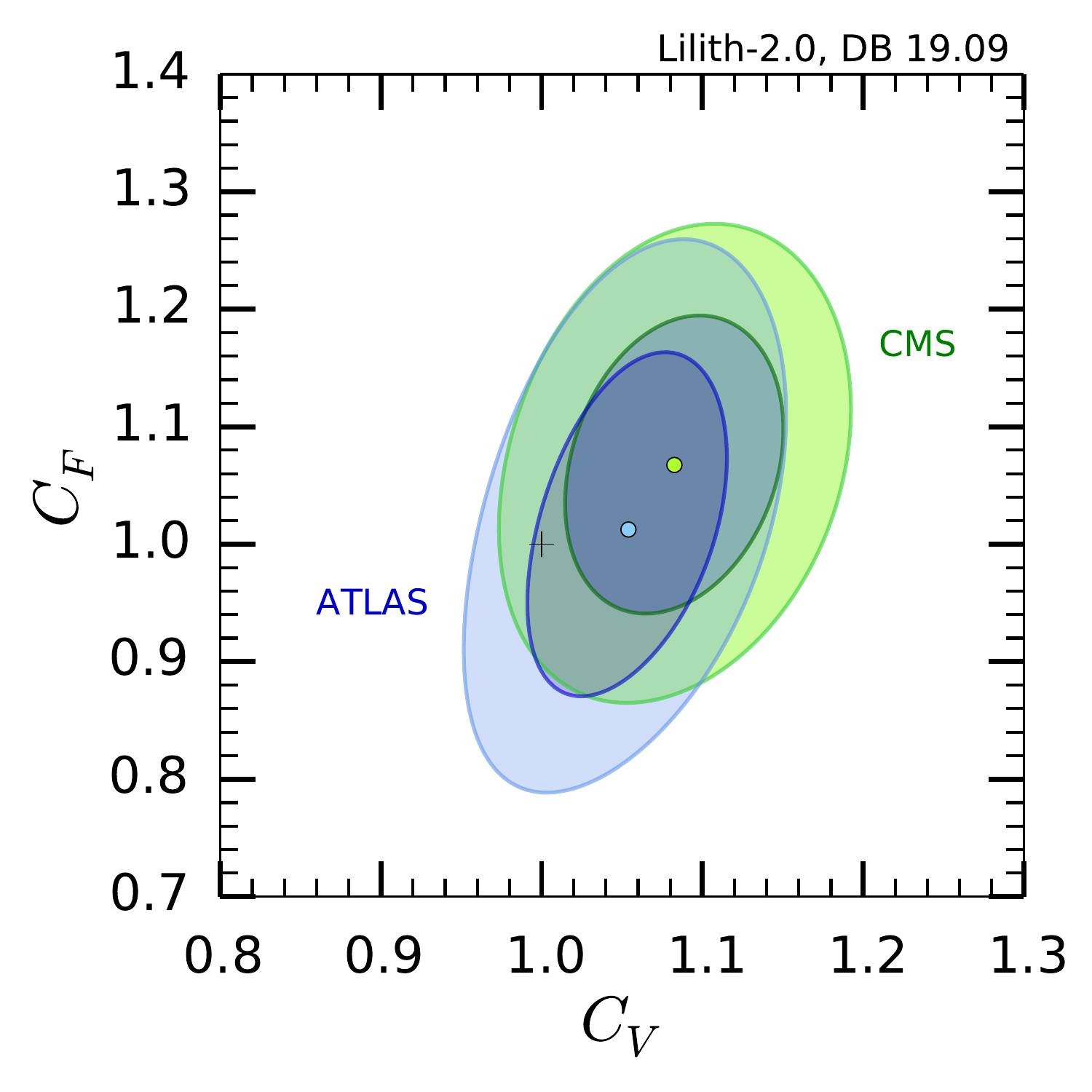}\quad%
\includegraphics[width=0.4\textwidth]{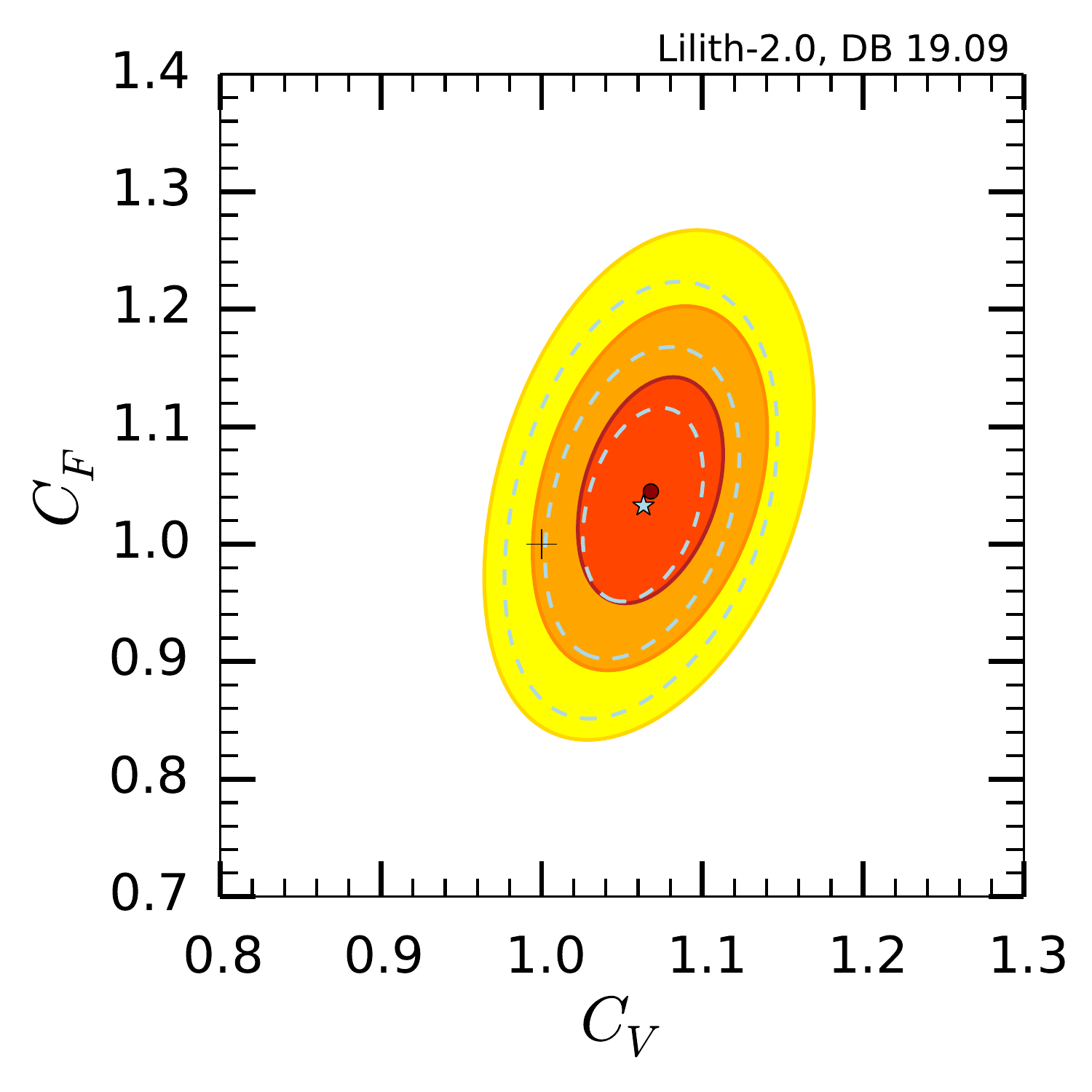}%
\caption{Fits of $C_F$ vs.\ $C_V$. Left: 68\% and 95\% CL regions for the Run~2 results from ATLAS (in blue) and from CMS (in green). 
Right: combination of the ATLAS and CMS Run~2 results in DB~19.09; the 68\%, 95\% and 99.7\% CL regions are shown as red, orange and yellow areas, respectively. 
In addition, the light-blue, dashed contours indicate the effect  when including also the Run~1 data.}
\label{fig:CVCF-fit}
\end{figure}

Let us now turn to the question how all these results taken together constrain the Higgs couplings. As an illustrative example, Figure~\ref{fig:CVCF-fit} presents combined fits of $C_F$ vs.\ $C_V$. The left panel shows the situation when using either the ATLAS (in blue) or the CMS (in green) Run~2 results in DB~19.09. 
As can be seen, the two experiments agree at the level of about $1\sigma$, 
the ATLAS results being slightly closer to the SM (marked as a black $+$). The situation when combining the results from both experiments is shown in the right panel.

\begin{figure}[t!]\centering
\includegraphics[width=0.43\textwidth]{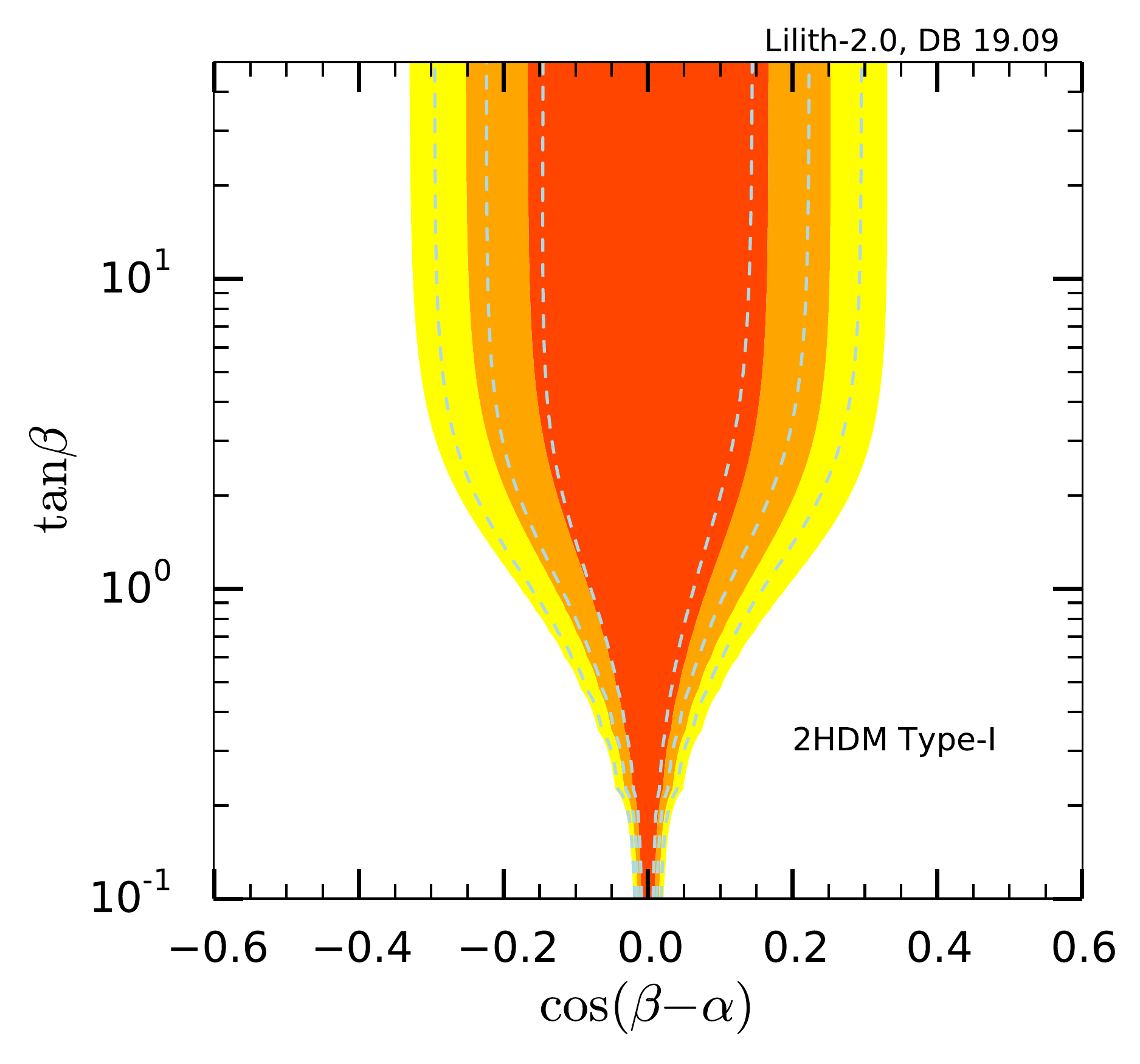}%
\includegraphics[width=0.43\textwidth]{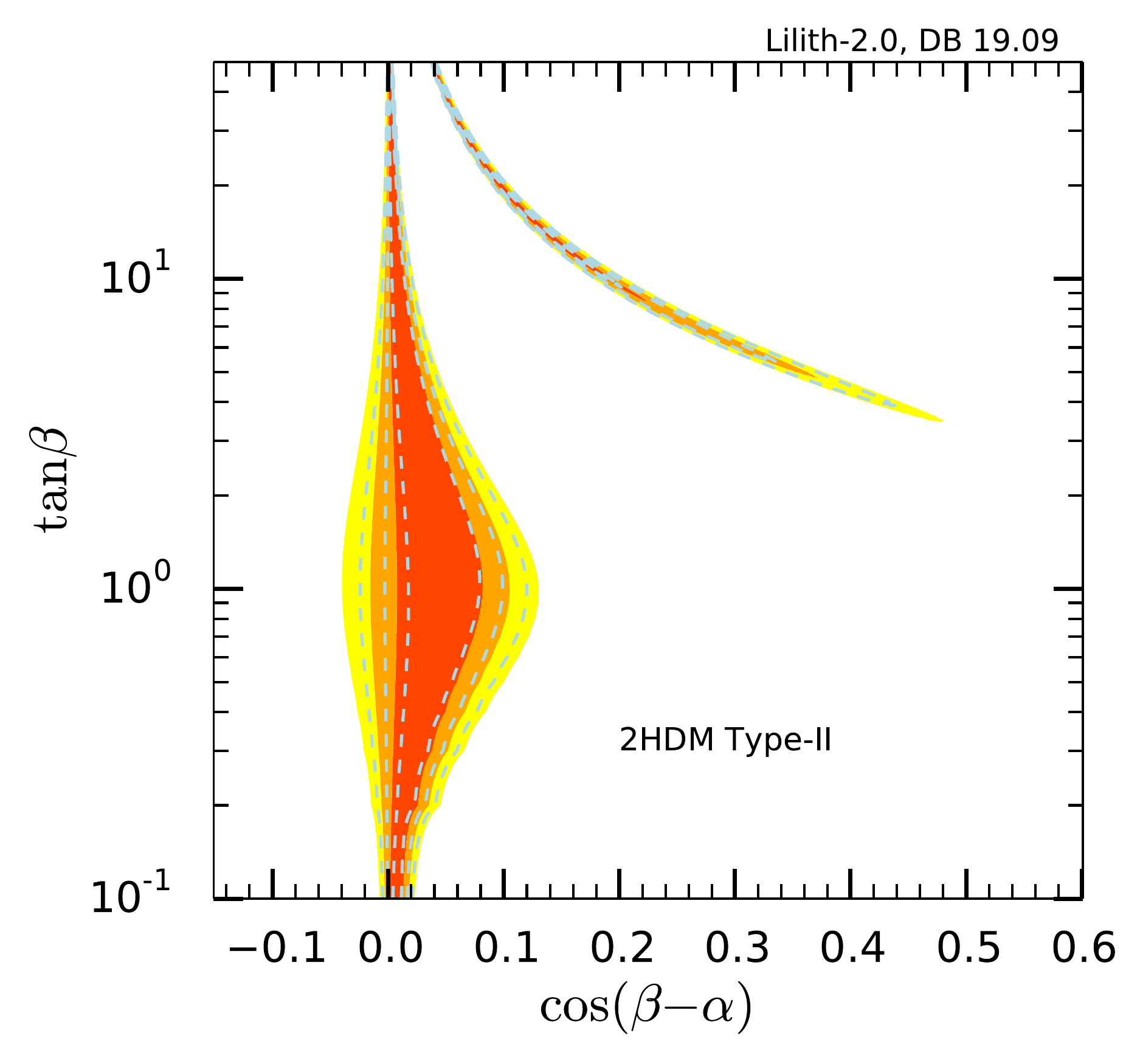}%
%\vspace*{-2mm}
\caption{Fits of $\tan\beta$ vs.\ $\cos(\beta-\alpha)$ for the 2HDM of Type~I (left) and of Type~II (right) 
from a combination of the ATLAS and CMS Run~2 results in DB~19.09. 
The red, orange and yellow areas are the 68\%, 95\% and 99.7\% CL regions, respectively. 
In addition, the light-blue, dashed contours indicate the 68\%, 95\% and 99.7\% CL regions when combining the Run~2 and Run~1 data. 
Loop contributions from charged Higgs bosons are neglected and decays into non-SM particles (such as $h\to AA$) assumed to be absent.}
\label{fig:2hdm-fit}
\end{figure}

As an example for a model with extended Higgs sector we consider  the 2HDM of Type~I and Type~II. 
The couplings of the lighter scalar $h$ are $C_F=\cos\alpha/\sin\beta$ in Type~I, and 
$C_U=\cos\alpha/\sin\beta$ and $C_D=-\sin\alpha/\cos\beta$ in Type~II; $C_V=\sin(\beta-\alpha)$ in both models. 
The fit results in the $\tan\beta$ vs.\ $\cos(\beta-\alpha)$ plane are shown in Figure~\ref{fig:2hdm-fit}. 
Note that for Type~II the banana-shaped second branch corresponds to the ``opposite-sign'' solution 
for the bottom Yukawa coupling~\cite{Ferreira:2014naa}.

\begin{figure}[t!]\centering
\includegraphics[width=0.45\textwidth]{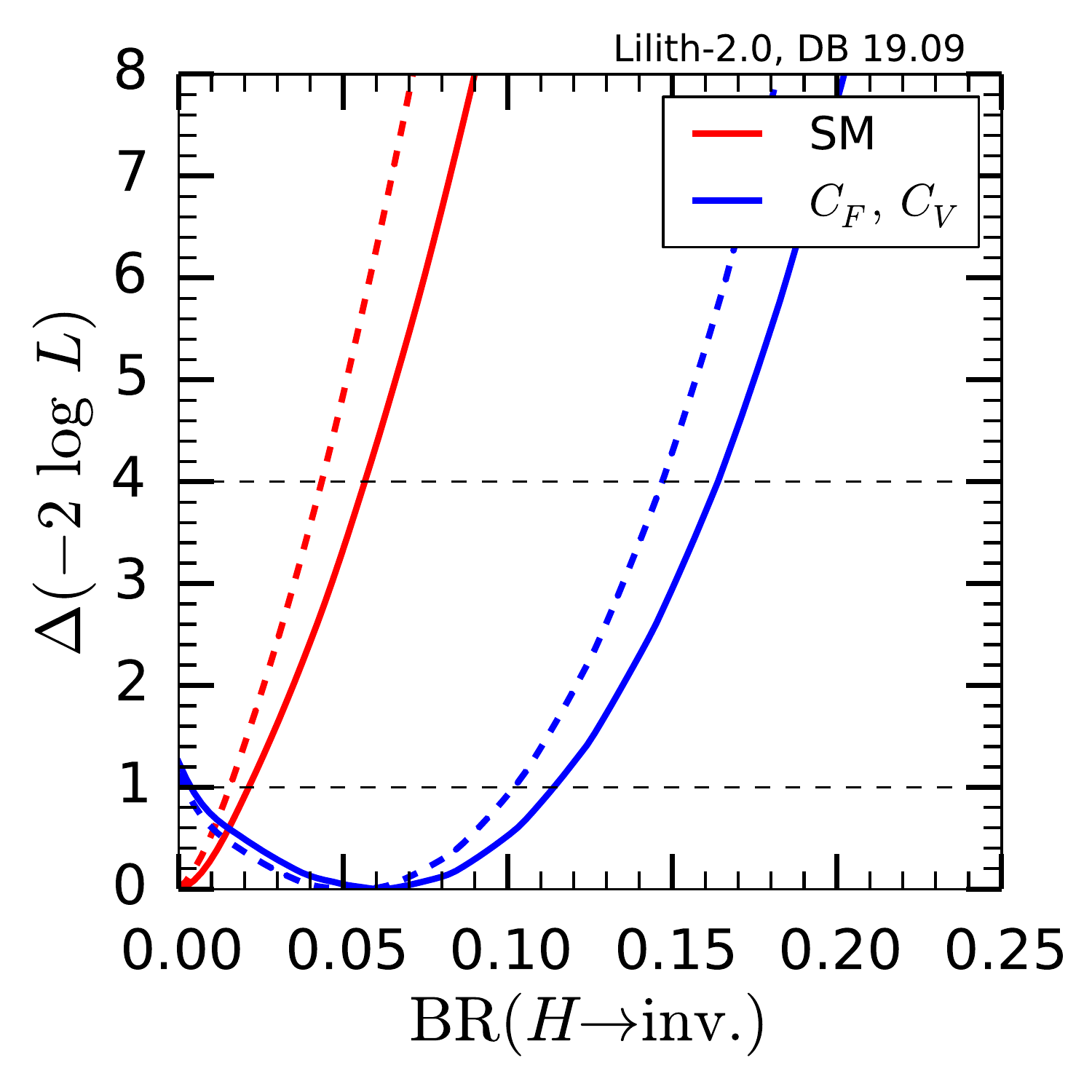}%
%\vspace*{-2mm}
\caption{1D profile likelihoods of BR($H\to {\rm inv.}$), in red for SM ($C_F=C_V=1$) couplings, 
in blue for $C_F$, $C_V$ as free parameters; full lines are for Run~2, while dashed lines are for Run~2 + Run~1 results in DB~19.09.}
\label{fig:BRinv}
\end{figure}

The signal strength measurements in the SM final states also strongly constrain possible invisible (or undetected) Higgs decays, $H\to {\rm inv.}$ (or $H\to {\rm undet.}$), as they would reduce the rates of (visible) SM final states~\cite{Belanger:2013kya}, see Eq.~\eqref{eq:BRinv}.    
For illustration, Figure~\ref{fig:BRinv} shows the 1D profile likelihood of 
${\rm BR}(H\to {\rm inv.})$ for two cases, SM couplings (in red) and $C_F$ and $C_V$ as free parameters (in blue). 
We find that the Run~2 (Run~2+Run~1) results in DB~19.09 constrain ${\rm BR}(H\to {\rm inv.})\lesssim 5\%$ (4\%) at 95\%~CL 
for the SM-like case,
and to ${\rm BR}(H\to {\rm inv.})\lesssim 16\%$ (15\%) when $C_F$ and $C_V$ are treated as free parameters; 
the case of free $C_U$, $C_D$, $C_V$ is not shown but gives the same result.   These constraints are much stronger than those from the dedicated searches for $H\to {\rm inv.}$ decays.

%%%%%%%%%%%%%%%%%%%%%%%%%%%%%%%%%%%%%%%%%%%%%%%%%%%%%%
\section{Ongoing developments and deployment}
%%%%%%%%%%%%%%%%%%%%%%%%%%%%%%%%%%%%%%%%%%%%%%%%%%%%%%

{\tt Lilith} is open source software distributed under the terms of the GNU General Public License. It is available from 
\begin{quote}
   \url{https://github.com/sabinekraml/Lilith-2/}. 
\end{quote}
The current official release, v2.0 with DB19.09 (see the \href{https://github.com/sabinekraml/Lilith-2/releases}{releases page}) is written in Python\,2.  
A Python\,3 version is available as v2.1 pre-release. The usage of the code is explained in \cite{Bernon:2015hsa} and the \href{https://indico.cern.ch/event/955391/contributions/4086275/}{tutorial session} of this workshop.  
%For the time being, it can be obtained from the \href{https://github.com/sabinekraml/Lilith-2/tree/py3-fullRun2}{py3-fullRun2} branch of the github repository. \textcolor{red}{\it [turn into v2.1 (pre-)release]}

\begin{figure}[t!]\centering
\includegraphics[width=0.5\textwidth]{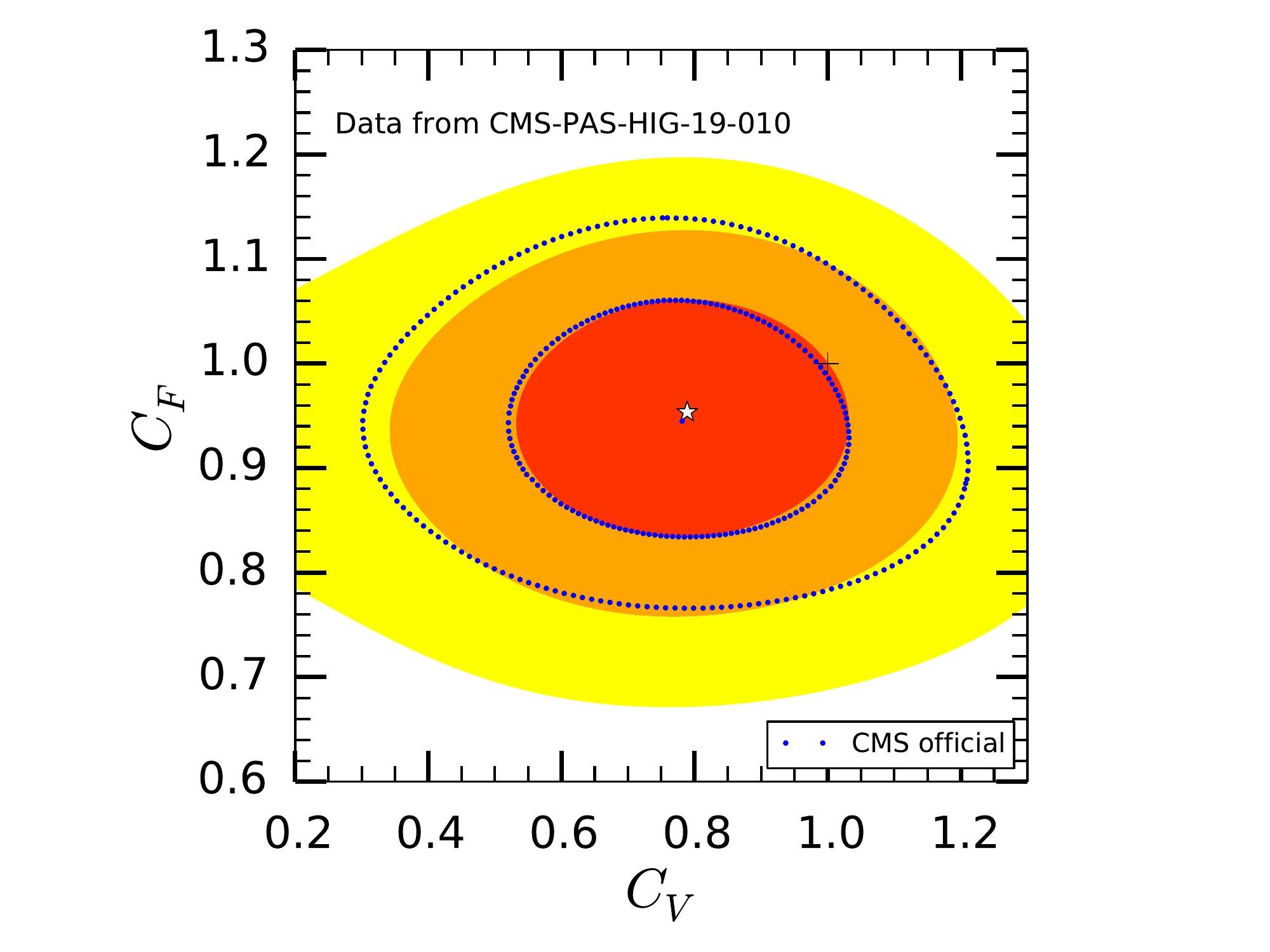}\hspace*{-1cm}
\includegraphics[width=0.5\textwidth]{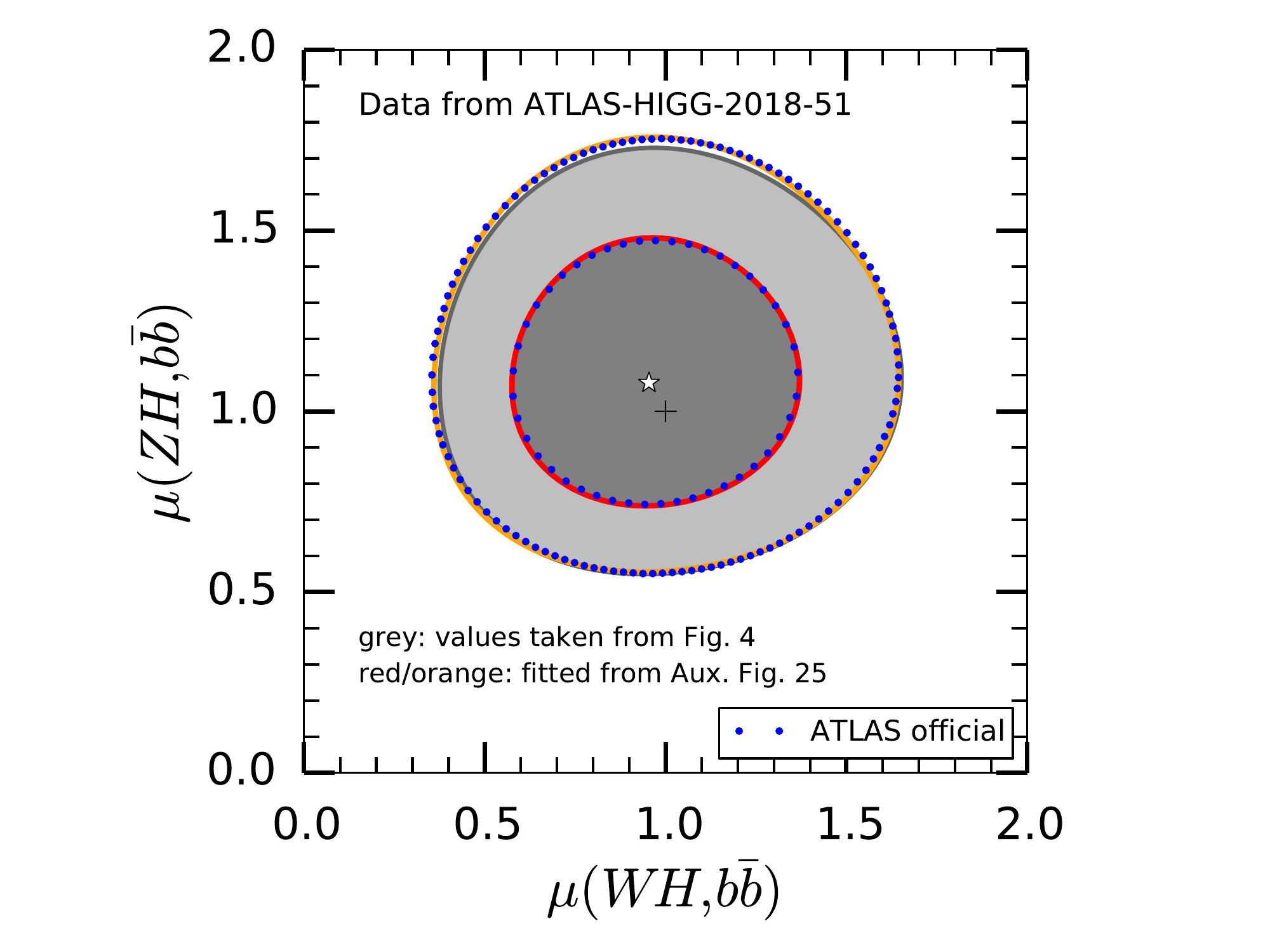}%
\caption{Validation plots for the implementation of the results of the CMS-PAS-HIG-19-010~\cite{CMS-PAS-HIG-19-010} (left) and ATLAS-HIGG-2018-51~\cite{Aad:2020jym} (right) analyses; see text for details.}
\label{fig:examples140fbi}
\end{figure}

Ongoing developments concentrate on the extension of the database with the ATLAS and CMS results for full Run~2 luminosity. 
Some implementations+validations are straightforward, like for the CMS measurement in the $H\to\tau\tau$ channel,   CMS-PAS-HIG-19-010~\cite{CMS-PAS-HIG-19-010}, shown in Figure~\ref{fig:examples140fbi} (left): 
here, the best-fit values and uncertainties for $\mu({\rm VBF}, \tau\tau)$ and  $\mu({\rm ggH}, \tau\tau)$ are taken from Fig.~10 of  \cite{CMS-PAS-HIG-19-010} and their
correlation $\rho=-0.265$ fitted from the 95\% CL contour in Fig.~14-b of \cite{CMS-PAS-HIG-19-010}. %(The uncertainties extracted from the 95\% CL contour of Fig.~14-b agree with those reported in Fig.~10.) 
The likelihood is modelled as a 2D variable Gaussian. For validation, we compare to the 68\% and 95\% CL contours in the $C_F$ vs.\ $C_V$ plane shown in Fig.~14-a of the CMS paper and find good agreement.  

For other publications, like the ATLAS measurements of $WH$ and $ZH$ production in the $H\to b\bar b$ decay channel, ATLAS-HIGG-2018-51~\cite{Aad:2020jym}, the possibilities for validation are limited to reproducing contours of constant signal strengths, see Figure~\ref{fig:examples140fbi} (right).\footnote{The results of this ATLAS paper are available on HEPData, which facilitates a lot their reuse.}  
Here, the grey contours show the 68\% and 95\% CL regions obtained when taking the  uncertainties 
for $\mu(WH, b\bar b)$ and $\mu(ZH, b\bar b)$ from Fig.~4 of~\cite{Aad:2020jym}, with 
correlation $\rho=0.027$ as mentioned in the paper's text. The comparison with the official ATLAS contours from Auxiliary Fig.~25 available on the analysis' twiki page (blue dots) allows to validate the shape of the likelihood, again approximated as a variable Gaussian. The agreement can be improved by fitting the uncertainties on  $\mu(WH, b\bar b)$ and $\mu(ZH, b\bar b)$ directly from the 95\% CL contour of  Aux.\ Fig.~25, as illustrated by the red and orange contours. 

There are also a number of new ATLAS and CMS measurements, which do not provide any validation material at all, in which case the (level of accuracy of the) implementations in {\tt Lilith} cannot be verified.   
New difficulties also arise from the fact that some ATLAS studies \cite{Aad:2020mkp,ATLAS-CONF-2020-027} report only simplified template cross sections (STXS) with SM uncertainties factored out, but not any more the per-channel signal strengths including SM uncertainties. 
The differences in the ATLAS and CMS policies in this respect were discussed by Davide Mungo at the Higgs\,2020 conference~\cite{mungo:higgs2020}. 
We are currently working on the procedure to fold the SM theory uncertainties back in, in order to reproduce the official ATLAS fits with good precision. %This is, however, not trivial and introduces new systematic uncertainties in the {\tt Lilith} results. 

%%%%%%%%%%%%%%%%%%%%%%%%%%%%%%%%%%%%%%%%%%%%%%%%%%%%%%
\section{Conclusions}
%%%%%%%%%%%%%%%%%%%%%%%%%%%%%%%%%%%%%%%%%%%%%%%%%%%%%%

We presented {\tt Lilith-2}, a light and easy-to-use Python tool for constraining new physics from signal strength measurements of the 125 GeV Higgs boson. The main novelties of v2.0 are:  
\begin{itemize}
\item a better treatment of asymmetric uncertainties through the use of variable Gaussian or Poisson likelihoods where appropriate; 
\item the use of multi-dimensional correlations whenever available;
\item a well-validated database (DB~19.09) including the published ATLAS and CMS Run~2 Higgs results for 36~fb$^{-1}$.
\end{itemize}
We discussed the modelling of the likelihood and the implementation and validation of the results in the database. Moreover, we showed examples for combined fits of reduced Higgs couplings, 2HDMs of Type~I and Type~II, and invisible Higgs decays. 
 
{\tt Lilith-2} is publicly available on \href{https://github.com/sabinekraml/Lilith-2}{GitHub} and ready to be used to constrain a wide class of new physics scenarios. 
It is also interfaced from {\tt micrOMEGAs}~\cite{Barducci:2016pcb} (v4.3 or higher) and can thus easily be used for dark matter models---this is particularly useful for, e.g., constraining Higgs decays into dark matter particles in a consistent manner.  
The code was recently translated from Python~2 to Python~3 in order to meet up-to-date standards. 
An extension of the database with the available ATLAS and CMS results for full Run~2 luminosity (about 140~fb$^{-1}$) is in preparation.

Given the high interest and ease of use of the signal strength framework---indeed a multitude of theory studies are based on Higgs signal strength results---we kindly ask the experimental collaborations 
to continue to provide detailed $\mu$ values for the various Higgs production$\times$decay modes,  
including their correlations. %, in addition to STXS and differential fiducial cross-sections. 
Reference~\cite{Abdallah:2020pec} lists detailed recommendations in this respect, to achieve optimal usefulness of the experimental results. 

Last but not least we want to note that the publication of full statistical likelihoods, as pioneered in~\cite{ATL-PHYS-PUB-2019-029} for supersymmetry  searches, would be extremely welcome also for Higgs results: it would render the various approximations for modelling the shape of the likelihood (and the associated try-and-error approach in the validation) unnecessary and thus  constitute a \emph{major leap} forward regarding the precision and reliability with which the Higgs results can be reused.

%===================================================================================
\section*{Acknowledgements}
%===================================================================================

%The work of M.B.\ and S.K.\
This work is supported in part by the IN2P3 master project ``Th\'eorie -- BSMGA''. 
D.T.N.\ and L.D.N.\ are funded by the Vietnam
National Foundation for Science and Technology Development (NAFOSTED) under grant number 103.01-2020.17.  

%%%%%%%%%%%%%%%%%%%%%%%%%%%%%%%%%%%%%%%%%%%%%%%%%%%%%%
%\bibliographystyle{JHEP}
%\bibliography{references.bib}

\begin{thebibliography}{10}

\bibitem{Bernon:2014vta}
J.~Bernon, B.~Dumont and S.~Kraml, \emph{{Status of Higgs couplings after run 1
  of the LHC}}, \href{https://doi.org/10.1103/PhysRevD.90.071301}{\emph{Phys.
  Rev.} {\bfseries D90} (2014) 071301}
  [\href{https://arxiv.org/abs/1409.1588}{{\ttfamily 1409.1588}}].

\bibitem{Bernon:2015hsa}
J.~Bernon and B.~Dumont, \emph{{Lilith: a tool for constraining new physics
  from Higgs measurements}},
  \href{https://doi.org/10.1140/epjc/s10052-015-3645-9}{\emph{Eur. Phys. J.}
  {\bfseries C75} (2015) 440}
  [\href{https://arxiv.org/abs/1502.04138}{{\ttfamily 1502.04138}}].

\bibitem{Kraml:2019sis}
S.~Kraml, T.Q.~Loc, D.T.~Nhung and L.D.~Ninh, \emph{{Constraining new physics
  from Higgs measurements with Lilith: update to LHC Run 2 results}},
  \href{https://doi.org/10.21468/SciPostPhys.7.4.052}{\emph{SciPost Phys.}
  {\bfseries 7} (2019) 052} [\href{https://arxiv.org/abs/1908.03952}{{\ttfamily
  1908.03952}}].

\bibitem{Boudjema:2013qla}
F.~Boudjema et~al., \emph{{On the presentation of the LHC Higgs Results}},  in
  \emph{{Workshop on Likelihoods for the LHC Searches Geneva, Switzerland,
  January 21-23, 2013}}, 2013
  [\href{https://arxiv.org/abs/1307.5865}{{\ttfamily 1307.5865}}].

\bibitem{Belanger:2013xza}
G.~Belanger, B.~Dumont, U.~Ellwanger, J.F.~Gunion and S.~Kraml, \emph{{Global
  fit to Higgs signal strengths and couplings and implications for extended
  Higgs sectors}},
  \href{https://doi.org/10.1103/PhysRevD.88.075008}{\emph{Phys. Rev.}
  {\bfseries D88} (2013) 075008}
  [\href{https://arxiv.org/abs/1306.2941}{{\ttfamily 1306.2941}}].

\bibitem{Belanger:2012gc}
G.~Belanger, B.~Dumont, U.~Ellwanger, J.F.~Gunion and S.~Kraml, \emph{{Higgs
  Couplings at the End of 2012}},
  \href{https://doi.org/10.1007/JHEP02(2013)053}{\emph{JHEP} {\bfseries 02}
  (2013) 053} [\href{https://arxiv.org/abs/1212.5244}{{\ttfamily 1212.5244}}].

\bibitem{Heinemeyer:2013tqa}
{\scshape LHC Higgs Cross Section Working Group} collaboration, \emph{{Handbook
  of LHC Higgs Cross Sections: 3. Higgs Properties}},
  \href{https://arxiv.org/abs/1307.1347}{{\ttfamily 1307.1347}}.

\bibitem{Barlow:2004wg}
R.~Barlow, \emph{{Asymmetric statistical errors}},  in \emph{{Statistical
  Problems in Particle Physics, Astrophysics and Cosmology (PHYSTAT 05):
  Proceedings, Oxford, UK, September 12-15, 2005}}, pp.~56--59, 2004
  [\href{https://arxiv.org/abs/physics/0406120}{{\ttfamily physics/0406120}}].

\bibitem{ATLAS-CONF-2020-045}
{\scshape ATLAS} collaboration, \emph{{Observation of vector-boson-fusion
  production of Higgs bosons in the $H \to WW^{\ast} \to e\nu\mu\nu$ decay
  channel in $pp$ collisions at $\sqrt{s}=13$ TeV with the ATLAS detector}},
  Tech. Rep.
  \href{https://atlas.web.cern.ch/Atlas/GROUPS/PHYSICS/CONFNOTES/ATLAS-CONF-2020-045/}{ATLAS-CONF-2020-045},
  CERN, Geneva (Aug, 2020).

\bibitem{Sirunyan:2018koj}
{\scshape CMS} collaboration, \emph{{Combined measurements of Higgs boson
  couplings in proton–proton collisions at $\sqrt{s}=13$~TeV}},
  \href{https://doi.org/10.1140/epjc/s10052-019-6909-y}{\emph{Eur. Phys. J.}
  {\bfseries C79} (2019) 421}
  [\href{https://arxiv.org/abs/1809.10733}{{\ttfamily 1809.10733}}].

\bibitem{Aaboud:2017vzb}
{\scshape ATLAS} collaboration, \emph{{Measurement of the Higgs boson coupling
  properties in the $H\rightarrow ZZ^{*} \rightarrow 4\ell$ decay channel at
  $\sqrt{s}$ = 13 TeV with the ATLAS detector}},
  \href{https://doi.org/10.1007/JHEP03(2018)095}{\emph{JHEP} {\bfseries 03}
  (2018) 095} [\href{https://arxiv.org/abs/1712.02304}{{\ttfamily
  1712.02304}}].

\bibitem{Aaboud:2018xdt}
{\scshape ATLAS} collaboration, \emph{{Measurements of Higgs boson properties
  in the diphoton decay channel with 36 fb$^{-1}$ of $pp$ collision data at
  $\sqrt{s} = 13$ TeV with the ATLAS detector}},
  \href{https://doi.org/10.1103/PhysRevD.98.052005}{\emph{Phys. Rev.}
  {\bfseries D98} (2018) 052005}
  [\href{https://arxiv.org/abs/1802.04146}{{\ttfamily 1802.04146}}].

\bibitem{Aaboud:2018jqu}
{\scshape ATLAS} collaboration, \emph{{Measurements of gluon-gluon fusion and
  vector-boson fusion Higgs boson production cross-sections in the $H \to
  WW^{\ast} \to e\nu\mu\nu$ decay channel in $pp$ collisions at $\sqrt{s}=13$
  TeV with the ATLAS detector}},
  \href{https://doi.org/10.1016/j.physletb.2018.11.064}{\emph{Phys. Lett.}
  {\bfseries B789} (2019) 508}
  [\href{https://arxiv.org/abs/1808.09054}{{\ttfamily 1808.09054}}].

\bibitem{Aaboud:2018pen}
{\scshape ATLAS} collaboration, \emph{{Cross-section measurements of the Higgs
  boson decaying into a pair of $\tau$-leptons in proton-proton collisions at
  $\sqrt{s}=13$ TeV with the ATLAS detector}},
  \href{https://doi.org/10.1103/PhysRevD.99.072001}{\emph{Phys. Rev.}
  {\bfseries D99} (2019) 072001}
  [\href{https://arxiv.org/abs/1811.08856}{{\ttfamily 1811.08856}}].

\bibitem{Aaboud:2017ojs}
{\scshape ATLAS} collaboration, \emph{{Search for the dimuon decay of the Higgs
  boson in $pp$ collisions at $\sqrt{s} = 13$~TeV with the ATLAS detector}},
  \href{https://doi.org/10.1103/PhysRevLett.119.051802}{\emph{Phys. Rev. Lett.}
  {\bfseries 119} (2017) 051802}
  [\href{https://arxiv.org/abs/1705.04582}{{\ttfamily 1705.04582}}].

\bibitem{Aaboud:2018gay}
{\scshape ATLAS} collaboration, \emph{{Search for Higgs bosons produced via
  vector-boson fusion and decaying into bottom quark pairs in $\sqrt{s} = 13$
  $\mathrm{TeV}$ $pp$ collisions with the ATLAS detector}},
  \href{https://doi.org/10.1103/PhysRevD.98.052003}{\emph{Phys. Rev.}
  {\bfseries D98} (2018) 052003}
  [\href{https://arxiv.org/abs/1807.08639}{{\ttfamily 1807.08639}}].

\bibitem{Aaboud:2019rtt}
{\scshape ATLAS} collaboration, \emph{{Combination of searches for invisible
  Higgs boson decays with the ATLAS experiment}},
  \href{https://doi.org/10.1103/PhysRevLett.122.231801}{\emph{Phys. Rev. Lett.}
  {\bfseries 122} (2019) 231801}
  [\href{https://arxiv.org/abs/1904.05105}{{\ttfamily 1904.05105}}].

\bibitem{Aad:2019lpq}
{\scshape ATLAS} collaboration, \emph{{Measurement of the production cross
  section for a Higgs boson in association with a vector boson in the $H
  \rightarrow WW^{\ast} \rightarrow \ell\nu\ell\nu$ channel in $pp$ collisions
  at $\sqrt{s} = 13$~TeV with the ATLAS detector}},
  \href{https://arxiv.org/abs/1903.10052}{{\ttfamily 1903.10052}}.

\bibitem{Aaboud:2017xsd}
{\scshape ATLAS} collaboration, \emph{{Evidence for the $ H\to b\overline{b} $
  decay with the ATLAS detector}},
  \href{https://doi.org/10.1007/JHEP12(2017)024}{\emph{JHEP} {\bfseries 12}
  (2017) 024} [\href{https://arxiv.org/abs/1708.03299}{{\ttfamily
  1708.03299}}].

\bibitem{Aaboud:2017bja}
{\scshape ATLAS} collaboration, \emph{{Search for an invisibly decaying Higgs
  boson or dark matter candidates produced in association with a $Z$ boson in
  $pp$ collisions at $\sqrt{s} =13$ TeV with the ATLAS detector}},
  \href{https://doi.org/10.1016/j.physletb.2017.11.049}{\emph{Phys. Lett.}
  {\bfseries B776} (2018) 318}
  [\href{https://arxiv.org/abs/1708.09624}{{\ttfamily 1708.09624}}].

\bibitem{Aaboud:2017jvq}
{\scshape ATLAS} collaboration, \emph{{Evidence for the associated production
  of the Higgs boson and a top quark pair with the ATLAS detector}},
  \href{https://doi.org/10.1103/PhysRevD.97.072003}{\emph{Phys. Rev.}
  {\bfseries D97} (2018) 072003}
  [\href{https://arxiv.org/abs/1712.08891}{{\ttfamily 1712.08891}}].

\bibitem{Aaboud:2017rss}
{\scshape ATLAS} collaboration, \emph{{Search for the standard model Higgs
  boson produced in association with top quarks and decaying into a $b\bar{b}$
  pair in $pp$ collisions at $\sqrt{s}=13$ TeV with the ATLAS detector}},
  \href{https://doi.org/10.1103/PhysRevD.97.072016}{\emph{Phys. Rev.}
  {\bfseries D97} (2018) 072016}
  [\href{https://arxiv.org/abs/1712.08895}{{\ttfamily 1712.08895}}].

\bibitem{Sirunyan:2018owy}
{\scshape CMS} collaboration, \emph{{Search for invisible decays of a Higgs
  boson produced through vector boson fusion in proton-proton collisions at
  $\sqrt{s} =13$~TeV}},  \href{https://arxiv.org/abs/1809.05937}{{\ttfamily
  1809.05937}}.

\bibitem{Sirunyan:2018cpi}
{\scshape CMS} collaboration, \emph{{Search for the associated production of
  the Higgs boson and a vector boson in proton-proton collisions at
  $\sqrt{s}=13$~TeV via Higgs boson decays to $\tau$ leptons}},
  \href{https://doi.org/10.1007/JHEP06(2019)093}{\emph{JHEP} {\bfseries 06}
  (2019) 093} [\href{https://arxiv.org/abs/1809.03590}{{\ttfamily
  1809.03590}}].

\bibitem{Ferreira:2014naa}
P.M.~Ferreira, J.F.~Gunion, H.E.~Haber and R.~Santos, \emph{{Probing wrong-sign
  Yukawa couplings at the LHC and a future linear collider}},
  \href{https://doi.org/10.1103/PhysRevD.89.115003}{\emph{Phys. Rev.}
  {\bfseries D89} (2014) 115003}
  [\href{https://arxiv.org/abs/1403.4736}{{\ttfamily 1403.4736}}].

\bibitem{Belanger:2013kya}
G.~Belanger, B.~Dumont, U.~Ellwanger, J.F.~Gunion and S.~Kraml, \emph{{Status
  of invisible Higgs decays}},
  \href{https://doi.org/10.1016/j.physletb.2013.05.024}{\emph{Phys. Lett.}
  {\bfseries B723} (2013) 340}
  [\href{https://arxiv.org/abs/1302.5694}{{\ttfamily 1302.5694}}].

\bibitem{CMS-PAS-HIG-19-010}
{\scshape CMS} collaboration, \emph{{Measurement of Higgs boson production in
  the decay channel with a pair of $\tau$ leptons}},  Tech. Rep.
  \href{https://cds.cern.ch/record/2725590}{CMS-PAS-HIG-19-010}, CERN, Geneva
  (2020).

\bibitem{Aad:2020jym}
{\scshape ATLAS} collaboration, \emph{{Measurements of $WH$ and $ZH$ production
  in the $H \rightarrow b\bar{b}$ decay channel in $pp$ collisions at 13 TeV
  with the ATLAS detector}},
  \href{https://arxiv.org/abs/2007.02873}{{\ttfamily 2007.02873}}.

\bibitem{Aad:2020mkp}
{\scshape ATLAS} collaboration, \emph{{Higgs boson production cross-section
  measurements and their EFT interpretation in the $4\ell $ decay channel at
  $\sqrt{s}=$13~TeV with the ATLAS detector}},
  \href{https://doi.org/10.1140/epjc/s10052-020-8227-9}{\emph{Eur. Phys. J. C}
  {\bfseries 80} (2020) 957}
  [\href{https://arxiv.org/abs/2004.03447}{{\ttfamily 2004.03447}}].

\bibitem{ATLAS-CONF-2020-027}
{\scshape ATLAS} collaboration, \emph{{A combination of measurements of Higgs
  boson production and decay using up to $139$ fb$^{-1}$ of proton--proton
  collision data at $\sqrt{s}=$ 13 TeV collected with the ATLAS experiment}},
  Tech. Rep. \href{http://cds.cern.ch/record/2725733}{ATLAS-CONF-2020-027},
  CERN, Geneva (Aug, 2020).

\bibitem{mungo:higgs2020}
D.P.~Mungo, \emph{{Combination of Higgs measurements: $\kappa$ modifiers, STXS
  framework and MSSM interpretation}},  {Talk given at the Higgs\,2020
  conference, 26-30 Oct.\ 2020 (online),
  \url{https://indico.cern.ch/event/900384/contributions/4059280/}}.

\bibitem{Barducci:2016pcb}
D.~Barducci, G.~Belanger, J.~Bernon, F.~Boudjema, J.~Da~Silva, S.~Kraml et~al.,
  \emph{{Collider limits on new physics within micrOMEGAs\_4.3}},
  \href{https://doi.org/10.1016/j.cpc.2017.08.028}{\emph{Comput. Phys. Commun.}
  {\bfseries 222} (2018) 327}
  [\href{https://arxiv.org/abs/1606.03834}{{\ttfamily 1606.03834}}].

\bibitem{Abdallah:2020pec}
{\scshape LHC Reinterpretation Forum}, \emph{{Reinterpretation of
  LHC Results for New Physics: Status and Recommendations after Run 2}},
  \href{https://doi.org/10.21468/SciPostPhys.9.2.022}{\emph{SciPost Phys.}
  {\bfseries 9} (2020) 022} [\href{https://arxiv.org/abs/2003.07868}{{\ttfamily
  2003.07868}}].

\bibitem{ATL-PHYS-PUB-2019-029}
{ATLAS collaboration}, \emph{{Reproducing searches for new physics with the
  ATLAS experiment through publication of full statistical likelihoods}},
  Tech. Rep. ATL-PHYS-PUB-2019-029, CERN, Geneva (Aug, 2019).

\end{thebibliography}

\providecommand{\href}[2]{#2}\begingroup\raggedright\endgroup

%%%%%%%%%%%%%%%%%%%%%%%%%%%%%%%%%%%%%%%%%%%%%%%%%%%%%%

\end{document}